\documentclass[aps,prl,twocolumn,preprintnumbers,amsmath,amssymb,superscriptaddress]{revtex4-1}
\usepackage{graphicx}
\usepackage{subfigure}
\usepackage{epsfig}
\usepackage{dcolumn}
\usepackage{bm}
\usepackage{ulem}
\usepackage{color}
\usepackage{multirow}
\usepackage{slashed}

\def\be{\begin{equation}}       \def\ee{\end{equation}}
\def\bea{\begin{eqnarray}}      \def\eea{\end{eqnarray}}
\def\ba{\begin{array}}
\def\ea{\end{array}}
\def\bnum{\begin{enumerate} }
\def\enum{\end{enumerate}}

\def\nn{\nonumber}

\def\=>{\Rightarrow}
\def\>{\rightarrow}

\def\eye2{Fathbb{I}}

\def\Eq#1{Eq.~(\ref{#1})}

\renewcommand{\>}{\rangle}

\begin{document}

\title{Chiral Tricritical Point: A New Universality Class in Dirac Systems}

\author{Shuai Yin}
\affiliation{Institute for Advanced Study, Tsinghua University, Beijing 100084, China}

\author{Shao-Kai Jian}
\email{jsk14@mails.tsinghua.edu.cn}
\affiliation{Institute for Advanced Study, Tsinghua University, Beijing 100084, China}

\author{Hong Yao}
\email{yaohong@tsinghua.edu.cn}
\affiliation{Institute for Advanced Study, Tsinghua University, Beijing 100084, China}
\affiliation{State Key Laboratory of Low Dimensional Quantum Physics, Tsinghua University, Beijing 100084, China}
\affiliation{Collaborative Innovation Center of Quantum Matter, Beijing 100084, China}

\begin{abstract}
Tricriticality, as a sister of criticality, is a fundamental and absorbing issue in condensed-matter physics. It has been verified that the bosonic Wilson-Fisher universality class can be changed by gapless fermionic modes at criticality. However, the counterpart phenomena at tricriticality have rarely been explored. In this Letter, we study a model in which a tricritical Ising model is coupled to massless Dirac fermions. We find that the massless Dirac fermions result in the emergence of a new tricritical point, which we refer to as the {\it chiral} tricritical point (CTP), at the phase boundary between the Dirac semimetal and the charge-density wave insulator. From functional renormalization group analysis of the effective action, we obtain the critical behaviors of the CTP, which are qualitatively distinct from both the tricritical Ising universality and the chiral Ising universality. We further extend the calculations of the chiral tricritical behaviors of Ising spins to the case of Heisenberg spins. The experimental relevance of the CTP in two-dimensional Dirac semimetals is also discussed.
\end{abstract}
\date{\today}
\maketitle

{\it Introduction.---} Discovering and understanding universality classes is among the central issues in modern condensed-matter physics as the fundamental concept of universality class is crucial to classifying the infrared physics in many-body systems~\cite{sachdevbook,Sondhi-RMP}. The universality class is dictated by the renormalization group (RG) flow near the corresponding fixed point~\cite{Fradkin-book,Xiaogang-book}. According to the number of relevant directions of the RG flow, the singular behavior near the fixed point can be classified into criticality, tricriticality, and so on~\cite{Cardy-book}. In contrast to the criticality controlled by only one relevant direction, tricriticality often shows more intriguing properties, since it has two relevant directions~\cite{Cardy-book,lawrie1984}. Thus, tricriticality has aroused enormous theoretical as well as experimental studies in condensed-matter systems~\cite{shenker1985,vojta1999,chubukov2004,senthil2004,cenke2017, jian2017,lonzarich1994,lonzarich1997,hayden2004,kato2015,grosche2017}. For instance, the tricritical Ising fixed point in $1+1$ dimensions enjoys superconformal symmetry \cite{shenker1985}. The quantum tricritical point (TP) connecting the deconfined quantum critical point \cite{senthil2004} and first-order transition in two-dimensional spin systems exhibits $O(4)$ symmetry and self-duality \cite{cenke2017}. Furthermore, even supersymmetric quantum electrodynamics can emerge at the pair-density-wave quantum TP on the surface of a three-dimensional topological insulator \cite{jian2017}. Among them, some theoretical predictions \cite{vojta1999,chubukov2004} have been verified in experiments~\cite{lonzarich1994,lonzarich1997,hayden2004,grosche2017}.

On the other hand, quantum phase transitions in systems of interacting Dirac fermions have been investigated intensively in recent years~\cite{Fradkin-book,Xiaogang-book}. Besides their fundamental relevance in relativistic quantum field theories, such fermion systems also emerge in the low-energy sector of various condensed-matter systems~\cite{sachdevbook,Sondhi-RMP}. It was realized that gapless fermionic excitations at quantum criticality play an essential role in determining the critical behaviors. For instance, coupling massless Dirac fermions to the Ising quantum critical point results in a qualitatively different universality class, usually called the chiral Ising universality class (or the Gross-Neveu-Yukawa universality class) if the Ising field couples to the Dirac fermions as a chiral mass \cite{gross1974,kovner1993,herbut2017a,herbut2017b,knorr2016,li2013, LWang2014NJP,ZXLi2015NJP,ZXLi2015PRB,YFJiang2017PRB,Raghu2008PRL}. Recently, it was proposed in Ref. \cite{li2017} by two of us and collaborators that the presence of gapless fermionic modes can even change the type of phase transition from first-order to continuous, realizing the so-called fermion-induced quantum critical points (also see Refs.~\cite{jian2016a,jian2016b,herbut2017c}). However, for the more colorful tricriticality, the effects induced by the massless Dirac fermions have rarely been studied. Some questions then arise: To what extent is the tricriticality affected by the Dirac fermion? Will a new fixed point emerge? If so, how are we to characterize the critical properties in this new universality class?

\begin{figure}[t]
    \includegraphics[width=5cm]{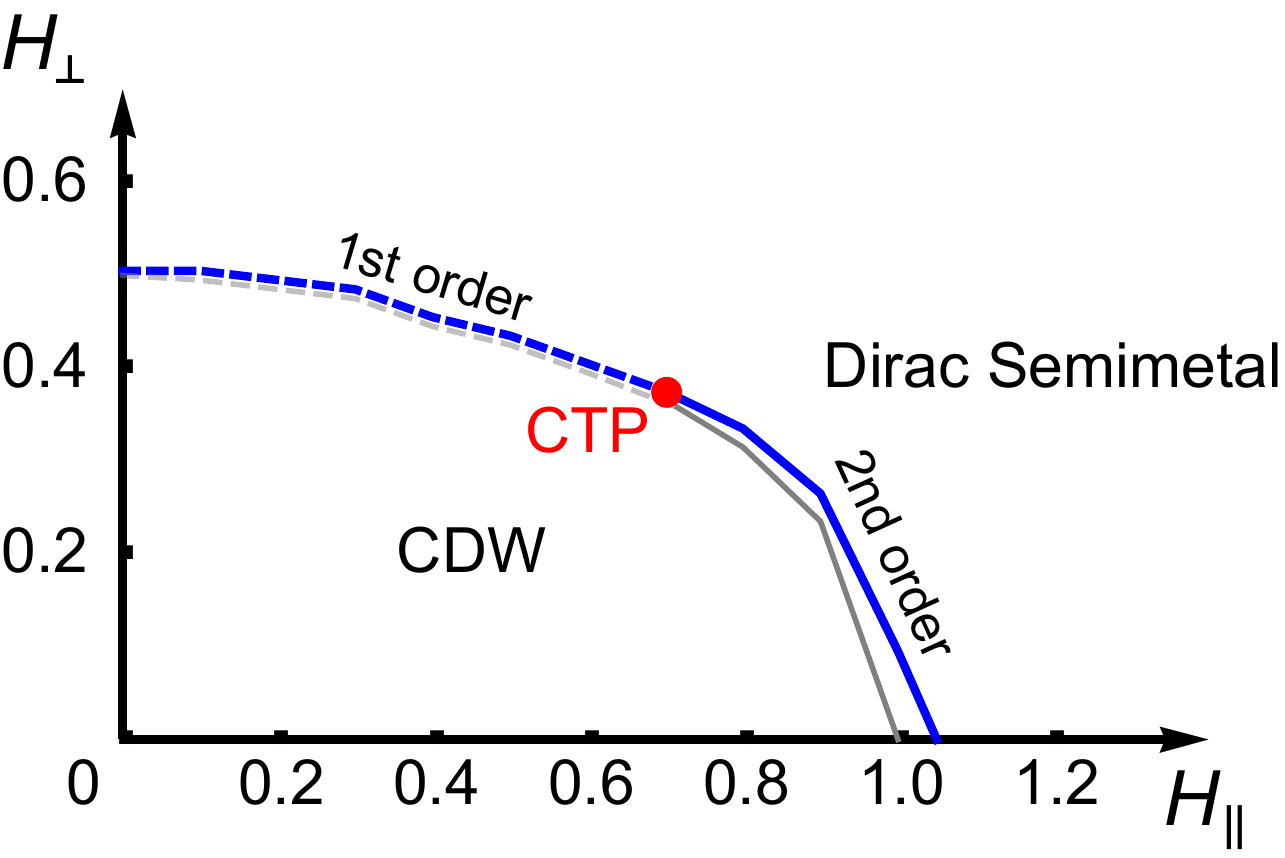}
\caption{\label{phase} The quantum phase diagram as a function of Zeeman fields $H_\perp$ and $H_{\parallel}$ with $h=0.2, t=1, J_1=2J_2=1/6$. The blue dashed and blue solid lines refer to first- and second-order phase transitions between the charge-density wave (CDW) insulator and the Dirac semimetal. The red point at $(H_\parallel, H_\perp)=(0.7,0.375)$ denotes the chiral tricritial point (CTP). The gray dashed and gray solid lines refer to first- and second-order phase transitions of the Ising field without coupling to fermions ($h=0$).}
\end{figure}

To answer these questions, in this Letter we report a first theoretical study of the chiral tricritical point (CTP). As suggested by its name, the CTP refers to a tricritical point coupled to massless Dirac fermions. More specifically, we first construct a microscopic model on the honeycomb lattice, where the Ising field couples to the itinerant fermions as a staggered on-site potential. Using mean-field (MF) calculations, we find that this microscopic model features a zero-temperature transition between a charge-density wave insulator (CDW) and a Dirac semimetal as shown in Fig.~\ref{phase}, where the transition can be either discontinuous or continuous depending on two tuning parameters. The CTP emerges at the location where first- and second-order transitions meet.

To understand the universal behaviors of the CTP, we carry out functional renormalization group (FRG) calculations of the effective theory near the transition point. By solving the RG equations, we find four nontrivial fixed points, corresponding to the Ising, chiral Ising, tricritical Ising, and chiral tricritical Ising universality classes, distinguished by the relevant directions. The former three universal classes have been extensively explored~\cite{Cardy-book, kovner1993, herbut2017b,vichi2012, lawrie1984}, while the last one--- i.e., the chiral tricritical Ising universality class---is new and is our main focus in the present Letter. Various critical exponents characterizing this novel universality class are obtained. We further study tricritical points of Heisenberg transitions in Dirac semimetals \cite{kovner1993, Herbut2006PRL,herbut2017b, herbut2014, knorr2017, herbut2013} by extending the Ising fields to Heisenberg spins. The relevance of these universality classes to real materials like graphene or graphene-like systems will also be discussed.

{\it Lattice model and phase diagram.---} It is known that the (2+1)-dimensional Ising model under both transversal and longitudinal Zeeman fields can realize a tricritical point at the phase boundary between first- and second-order transitions \cite{kato2015}. The Hamiltonian on the honeycomb lattice is given by
\bea
H_b \!=\! J_1 \sum_{\langle ij \rangle} \sigma_i^z \sigma_j^z \!-\! J_2 \sum_{\langle\langle ij \rangle\rangle} \sigma_i^z \sigma_j^z \!-\! H_\perp \sum_i \sigma_i^z \!-\! H_{\parallel} \sum_i \sigma_i^x,~~~~
\eea
where $\sigma$ denotes the Ising spins defined at each site, and $\langle$ $\rangle$ and $\langle\langle$ $\rangle\rangle$ refer to nearest neighbors (NN) and next-nearest neighbors (NNN), respectively. $J_1$ and $J_2$ refer to the NN antiferromagnetic and NNN ferromagnetic couplings, respectively. $H_\perp$ and $H_{\parallel}$ refer to the longitudinal and transversal Zeeman fields, respectively.

We assume the vacuum expectation value (VEV) of Ising variables is given by $\langle \sigma_{A/B}^z \rangle = m \pm \phi $, where $A$ and $B$ refer to two sublattices of the honeycomb lattice, and $m$ ($\phi$) denotes the uniform (staggered) VEV of the Ising variables. Note that we refer to microscopic degrees of freedom $\sigma$ as the Ising variable while referring to $\phi$ as the Ising field. By employing the MF decoupling, we obtain
\bea
&&\frac{H^\text{MF}_b}{M} = (J_- m- J_+ \phi - H_\perp) \sigma_A^z - H_{\parallel} \sigma_A^x \nn\\
&&~~~+ (J_- m + J_+ \phi - H_\perp) \sigma_B^z - H_{\parallel} \sigma_B^x - J_- m^2 + J_+ \phi^2,~~~~
\eea
where $J_\pm \equiv 3J_1 \pm 6 J_2$, and $M$ is the number of unit cells. Below, we focus on the case where $J_-=0$ for simplicity, because $m$ decouples from the MF Hamiltonian for $J_-=0$. Such a  simplification will not qualitatively change the physics discussed here, because the global $Z_2$ symmetry---namely, the symmetry generated by $\prod_i \sigma^x_i$---is already broken by the external longitudinal field $H_\perp$.

As a function of $H_\perp$ and $H_\parallel$, the Hamiltonian $H_b$ exhibits a phase transition from $\langle\phi\rangle=0$ to $\langle\phi\rangle\ne 0$ (breaking the lattice inversion or $Z_2$ symmetry that exchanges two sublattices), which is continuous or discontinuous depending on the external fields (see the Supplemental Material for details) and a quantum tricritical point separating these two types of transitions, as shown by the gray line in Fig.~\ref{phase}. The Landau free energy up to sixth order near the transitions is given by (see the Supplemental Material for details) $f_b=\sum_{n=0}^3 \lambda''_{2n} \phi^{2n}$, where $\lambda''_2 \propto (H_\perp^2 + H_{\parallel}^2)^{3/2}-H_{\parallel}^2 J_+$, $\lambda''_4\propto  H_{\parallel}^2-4  H_\perp^2$, and $\lambda''_6$ is a positive constant. The tricritical point is located at $\lambda_2=\lambda_4=0$, which can be accessed by tuning two parameters, $H_\perp$ and $H_\parallel$. Since the critical behaviors of this quantum tricritical point are governed by a $\phi^6$ theory whose upper critical dimension is 3, the MF theory should be reliable. Various critical exponents of this tricritical Ising universality class are listed in Table \ref{tabexp}.

To investigate a CTP, we consider massless Dirac fermions on the honeycomb lattice which couple to the Ising variables as follows:
\bea\label{Hf}
	H_f = -t \sum_{\langle ij\rangle, s} (c_{is}^\dag c_{js} + H.c.)+ h \sum_{i,s} \sigma_i^z c_{is}^\dag c_{is}, ~~~
\eea
where $c_{is}^\dag$ labels the operator creating a fermion on site $i$ with spin polarization $s$, $t$ is the hopping amplitude, and $h$ refers to the coupling strength between fermions and Ising variables. Note that in \Eq{Hf} there is an implicit chemical potential term $-\mu\sum_{i,s}c_{is}^\dag c_{is}$. The condensation of the Ising field $\phi$ gives rise to staggered on-site potentials for the fermions on the honeycomb lattice which can gap out the Dirac fermions, as shown by the dispersions of the fermions in the Ising symmetry-breaking phase: $
\varepsilon(\vec k)= \pm \{h^2 \phi^2+ t^2 [3+2(\cos k_1+ \cos k_2+ \cos(k_1-k_2))]\}^{\frac12}$,
where the lattice constant is set to be 1 for simplicity, and $\hat e_1=(1,0)$, $\hat e_2=(\frac12,\frac{\sqrt{3}}2)$, and $k_i= \hat e_i \cdot \vec k$.

\begin{figure}
\subfigure[]{\includegraphics[width=3.cm]{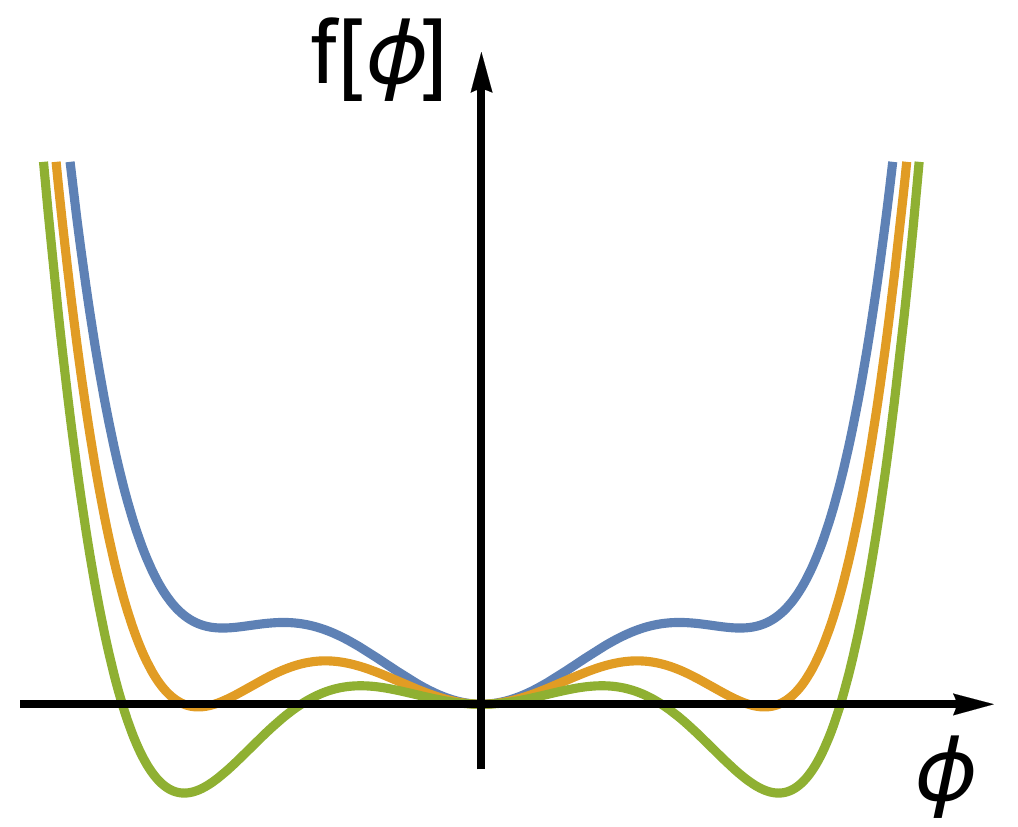}}~~~~
\subfigure[]{\includegraphics[width=3.1cm]{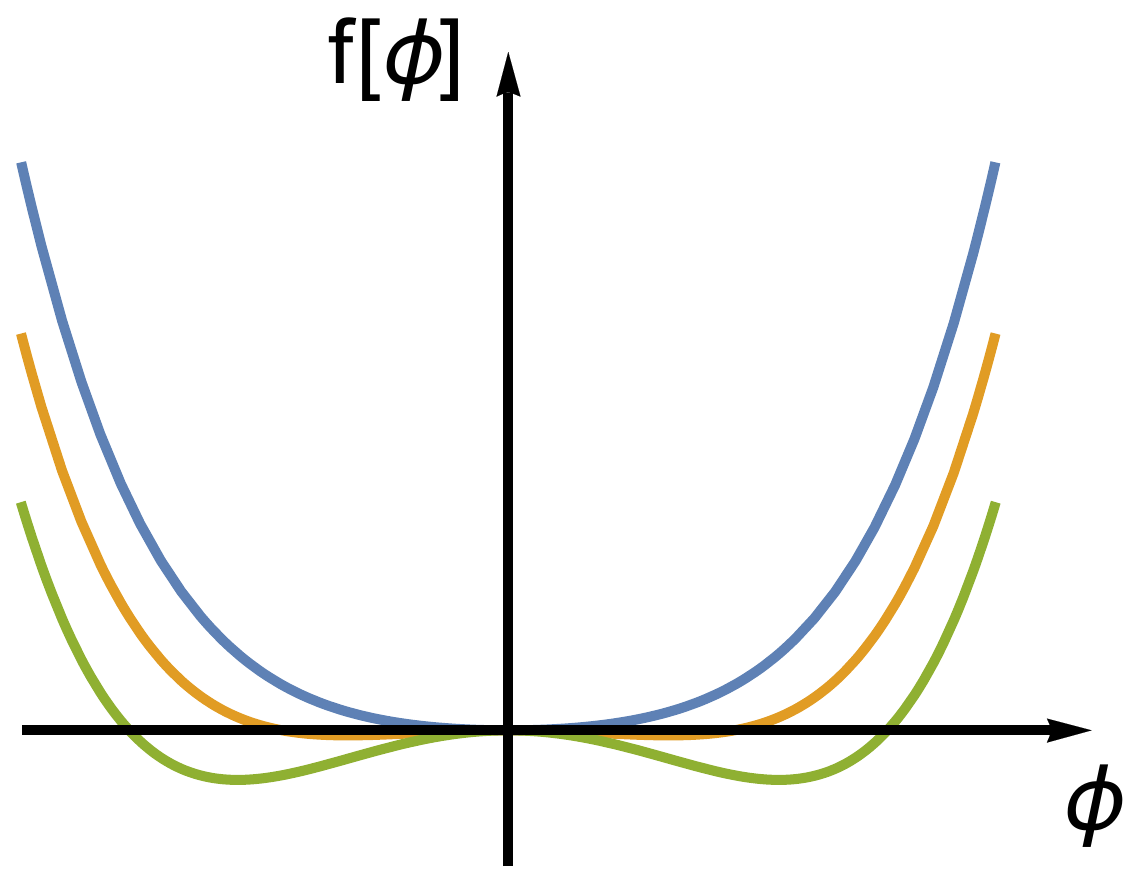}}
\caption{\label{free_energy} The evolution of the free energy as a function of the expectation value of the Ising field across the transition point. (a) The evolution of free energy shows a first-order transition at $H_\parallel=0.6$ as one tunes $H_\perp$. (b) The evolution of free energy shows a second-order transition at $H_\parallel=0.8$ as one tunes $H_\perp$. These provide the evidence that there is a quantum tricritical point between $0.6<H_\parallel<0.8$.}
\end{figure}

The quantum phase diagram of the Hamiltonian $H=H_b + H_f$ is shown in Fig.~\ref{phase}, where the blue line refers to the phase boundary, and the CTP is indicated by the red point (see the Supplemental Material for details). The evolutions of the free energy as a function of the expectation value of the Ising field across the transition point are shown in Fig. \ref{free_energy}, where one can clearly see that the phase transition changes from first to second order as the tuning parameters cross the CTP.

Near the phase boundary, the free energy can be expanded as (see the Supplemental Material for details) $f=\sum_{n=0}^3 \lambda_{2n}' \phi^{2n}+ \lambda_3 |\phi|^3$, where $\lambda_{2n}'$ are modified coupling constants owing to the presence of Dirac fermions. These modifications only renormalize the location of the phase boundary. However, a new nonanalytical term $\lambda_3 |\phi|^3$ appears due to the IR singularity of the gapless fermionic modes, where $\lambda_3 = \frac{4 h^3}{9\pi t^2}$, and changes the universal critical behaviors in the IR. Since the gapless modes involve both fermions and bosons, it is important to keep the massless fermionic modes at the critical point. Thus, to explore the critical behaviors, we use the effective Lagrangian given as follows (see the Supplemental Material for details):
\bea
	\mathcal{L} &=& \bar \psi \gamma^\mu \partial_\mu  \psi+ \frac12 (\partial_\mu \phi)^2+ h \phi \bar \psi  \psi +  U(\rho),
\eea
where $\psi$ refers to the Dirac fermions, while $\phi$ denotes the $Z_2$ order parameter. $\bar \psi = \psi^\dag \gamma^0 $, $\gamma^0= \tau^z$, $\gamma^1= \tau^y \mu^z$, $\gamma^2= \tau^x$, and $\partial_\mu= (\partial_\tau, \vec \nabla)$. $\tau^i$ and $\mu^i$ are Pauli matrices referring to sublattice and valley degrees of freedom, respectively. We set both fermion and boson velocity to 1 due to the emergent Lorentz symmetry \cite{jian2015, herbut2016}. $U(\rho)$ is an analytical function of $\rho \equiv \phi^2/2$, representing the potential of the Ising field. Note that the potential is an even function of $\phi$ owing to the $Z_2$ symmetry of the model. The expansion of potential can be expressed as $U(\rho)= \sum_i \lambda_{2i} \rho^i $. In the following calculations, we set the number of the four-component Dirac fermions to be $N_f=2$, which corresponds to spin-1/2 fermions on the honeycomb lattice.

{\it FRG analysis.---} The RG flow of the effective action satisfies the Wetterich equation \cite{berges2002,metzner2012}
\bea
\partial_{\tau} \Gamma_k=\textrm{STr}[(\Gamma_k^{(2)}+R_k)^{-1}\partial_{\tau} R_k], \label{wett}
\eea
where $\tau=\textrm{ln} k$, $\Gamma_k$ is the effective action, $R_k$ is the regulator function, $\textrm{STr}$ denotes the supertrace, and $\Gamma_k^{(2)}$ reads
\bea
\Gamma_k^{(2)}(i,j)\equiv\frac{\overrightarrow{\delta}}{\delta\Phi_i}\Gamma_k\frac{\overleftarrow{\delta}}{\delta\Phi_j},
\eea
with $\Phi_i\equiv(\phi,\bar{\psi},\psi)$. We then define the dimensionless variables as
\bea
\tilde{\rho} &\equiv& Z_{\phi,k} k^{2-D} {\rho}, \quad \quad \quad \tilde{\psi}\equiv \sqrt{Z_{\psi,k}} k^{1-D} {\psi}, \\
\tilde{h}^2 &=& k^{D-4}Z_{\phi,k}^{-1}Z_{\psi,k}^{-2}h^2, \quad \tilde{U}(\tilde\rho) \equiv k^{-D} U({\rho}).
\eea

\begin{table*}[t]
\centering
\caption{Critical exponents of the chiral tricritical Ising and chiral tricritical Heisenberg universality classes for $N_f=2$. The critical exponents of tricritical Ising, chiral Ising, and chiral Heisenberg universality are also listed for comparison. Note that the mean field calculation is exact for tricritical Ising universality in $2+1$ dimensions.}
  \begin{tabular}{c |ccccccc}
  \hline
  \hline
  ~~~~~~~Universality class~~~~~~~ & ~~~~~~~$\alpha$~~~~~~~ &~~~~~~~$\beta$~~~~~~~ & ~~~~~~~$\gamma$~~~~~~~   & ~~~~~~~$\delta$~~~~~~~  & ~~~~~~~$\nu$~~~~~~~ & ~~~~~~~$\eta_b$~~~~~~~ & ~~~~~~~$\eta_f$~~~~~~~  \\
  \hline
  CTP Ising &0.694   &0.378 & 0.550  & 2.456  & 0.435 & 0.736 & 0.036 \\
  Tricritical Ising &1/2   &1/4 & 1  & 5  &1/2 & 0 &  \\
  Chiral Ising \cite{knorr2016}  & $-1.018$  &0.894 & 1.230  & 2.376 & 1.006 & 0.777 & 0.028  \\
  \hline
  \hline
  CTP Heisenberg &0.791  &0.404 & 0.401  & 1.994 & 0.403 & 1.004 & 0.102 \\
  Chiral Heisenberg \cite{knorr2017} &$-1.773$   &1.278 & 1.217  & 1.953 &1.257  & 1.032 & 0.071 \\
  \hline
  \end{tabular}
\label{tabexp}
\end{table*}

To solve Eq.~(\ref{wett}), we employ the modified local potential approximation (LPA'). According to LPA', one can obtain the $\beta$ functions for the expansion coefficients from
\bea
\partial_{\tau}\tilde{\lambda}_n=\frac{1}{n!}\frac{\partial^n}{\partial \tilde\rho^n}\partial_t\tilde{U}(\tilde\rho)|_{\tilde\rho=0}.
\eea
Similarly, one can obtain the $\beta$ function for $\tilde{h}^2$ according to
\bea
\partial_{\tau}\tilde{h}=\frac{-i}{N_f d_\gamma}\textrm{Tr}[\gamma_3\frac{\delta}{\delta \phi(0)}\frac{\overrightarrow{\delta}}{\delta \bar{\psi}(0)}\partial_t\Gamma_k \frac{\overleftarrow{\delta}}{\delta \psi(0)}]|_{\tilde\rho=0}.
\eea
Using the Litim cutoff regulators~\cite{litim2001}, i.e., $R_k=Z_{\phi,k}(k^2-q^2)\Theta(k^2-q^2)$ for the boson cutoff and $R_k=Z_{\psi,k}i\gamma_\mu q_\mu (k/q-1)\Theta(k^2-q^2)$ for the fermion cutoff where $\Theta$ is the step function, the flow equations read~\cite{berges2002}
\bea
&&\partial_{\tau} \tilde{\lambda}_2 = (\eta_{b,k}-2)\tilde{\lambda}_2-(1-\frac{\eta_{b,k}}{D+2}) \frac{\tilde{\lambda}_4}{(1+\tilde{\lambda}_2)^2} \frac{24 v_D}{D} \nonumber\\
&&~~\quad\quad +(1-\frac{\eta_{f,k}}{D+1}) h^2 d_\gamma N_f \frac{8 v_D}{D}, \label{beta1}\\
&&\partial_{\tau} \tilde{\lambda}_4 = (D-4+2\eta_{b,k}) \tilde{\lambda}_4 +\frac{\eta_{b,k}}{D+2} \frac{12 v_D}{D} \frac{12 \tilde{\lambda}_4^2}{(1+\tilde{\lambda}_2)^3} \nonumber\\
&&~~ -\frac{\eta_{b,k}}{D+2} \frac{12 v_D}{D} \frac{5 \tilde{\lambda}_6}{(1+\tilde{\lambda}_2)^2}-\frac{16 v_D}{D} d_\gamma N_f h^4 (1\!-\!\frac{\eta_f}{D+1}), ~~~~~\\
&&\partial_{\tau} \tilde{\lambda}_6 = (2D-6+3\eta_{b,k}) \tilde{\lambda}_6 -\frac{\eta_b}{D+2} \frac{144 v_D}{D} \frac{6 \tilde{\lambda}_4^3}{(1+\tilde{\lambda}_2)^4} \nonumber\\
&&~~+\frac{\eta_{b,k}}{D+2} \frac{144 v_D}{D} \frac{5 \tilde{\lambda}_4\tilde{\lambda}_6}{(1+\tilde{\lambda}_2)^3}-\frac{32 v_D}{D} d_\gamma N_f \tilde{h}^6 (1\!-\!\frac{\eta_{f,k}}{D+1}),  ~~~~~\\
&&\partial_{\tau} \tilde{h}^2 = (D-4+\eta_{b,k}+2 \eta_{f,k})\tilde{h}^2-\frac{16 v_D}{D}\frac{1}{1+\tilde{\lambda}_2}(1-\frac{\eta_{f,k}}{D+1}) \nn\\
&&~~\quad\quad-\frac{16 v_D}{D}\frac{1}{1+\tilde{\lambda}_2}(1-\frac{\eta_{b,k}}{D+2})\frac{1}{(1+\tilde{\lambda}_2)^2}, \label{beta4}
\eea
where $v_D^{-1}=2^{D+1}\pi^{D/2}\Gamma(D/2)$, $D$ is the spacetime dimension, and the running anomalous dimensions for boson and fermion fields, defined as $\eta_{b,k}\equiv-\partial_{\tau} Z_{\phi,k}/Z_{\phi,k}$ and $\eta_{f,k}\equiv-\partial_{\tau} Z_{\psi,k}/Z_{\psi,k}$, respectively, can be solved explicitly as
\bea
\eta_{b,k} &=&\frac{4 v_D}{D}2 N_f d_\gamma h^2 \frac{4-3D+2\eta_{f,k}}{8-4D}, \label{eta2}\\
\eta_{f,k} &=& \frac{8 v_D}{D} h^2 \frac{\eta_{b,k}}{1+D} \frac{1}{(1+\lambda_2)^2}.
\label{eta1}
\eea
For $k\rightarrow0$, the equations above give anomalous dimensions of the boson field and fermion field, respectively.

We then solve the flow equations in Eqs.~(\ref{beta1})-(\ref{beta4}) for the case of $N_f=2$ (namely spin-1/2 fermions). To identify the CTP fixed point, we resort to the stability matrix at each fixed point. The stability matrix is defined as $S_{i,j}=-\partial B_i/\partial \alpha_j$, in which $B\equiv (\partial_t \tilde\lambda_{2i},\partial_t \tilde{h}^2)$ and $\alpha\equiv (\tilde\lambda_{2i},\tilde{h}^2)$. As we mentioned above, Eqs.~(\ref{beta1})-(\ref{beta4}) admit four nontrivial fixed points. Among them, two show double positive eigenvalues of $S$ (see the Supplemental Material for details). One corresponds to the Ising critical point without the coupling to the massless fermion ($h^*=0$). Note that at the Ising critical point, $h$ is a relevant direction. The other is the CTP fixed point, at which $h^*\neq 0$. The remaining two fixed points correspond to the chiral Ising critical point and tricritical Ising fixed point with one and three relevant directions, respectively. From the eigenvalues of $S$ (See the Supplemental Material), we find that the coupling to the massless fermions changes the critical properties dramatically, i.e., the eigenvalues corresponding to the CTP are very distinct from those of the tricritcal Ising point.

We are ready to obtain the relevant directions in the CTP by solving the relevant eigenvectors of the positive eigenvalues. Two eigenvectors are (in the basis of $\alpha$) $v_1=(0.996, 0.086, 0.005, -0.015)$ and $v_2=(-0.016, 0.349, 0.022, 0.937)$. For $v_1$, the relevant direction is dominated by $\lambda_2$, while for $v_2$, the relevant direction is mainly lying in the $\lambda_4$-$h^2$ plane, and the weight in $h^2$ direction is larger. Therefore, in contrast to the usual tricritical point whose relevant directions are $\lambda_2$ and $\lambda_4$, the coupling to the massless Dirac fermions plays significant roles in the CTP.

This relevant direction can be understood heuristically as follows: Consider the free energy $f=\sum_{i=1}^3 \lambda_{2i} \phi^{2i} $ with $\lambda_4<0$ and $\lambda_6>0$. Without coupling to fermions, i.e., $h=0$,  the phase transition is discontinuous; two local potential minima appear at $\pm\phi_{\text{min}} \equiv \pm \sqrt{\frac{-(4+\sqrt{12})\lambda_2}{3\lambda_4}}$ at the transition point which locates at $\lambda_{2c}= \frac{\lambda_4^2}{4\lambda_6}$ where the three local minima are degenerate. The fourth-order term dominates near $\phi=\phi_\text{min}$. Now, turning on the boson-fermion coupling will result in a nonanalytical term $\lambda_3 |\phi|^3$, where $\lambda_3 \propto h^3/t^2$, with $t$ being the energy scale capturing the kinetic energy of fermions (the hopping amplitude in our lattice model). If $\lambda_3$ is sufficiently large, especially $\lambda_3 |\phi_\text{min}|^3 \gg \lambda_4 \phi_\text{min}^4 $ (equivalently, $h \gg t^{2/3} |\lambda_4|^{1/2} \lambda_6^{-1/6}$), the transition will be driven to continuous. As a result, between $h=0$ and $h\gg t^{2/3} |\lambda_4|^{1/2} \lambda_6^{-1/6}$, there must exist a tricritical point separating first- and second-order transitions. This also explains why besides $\lambda_2$, the other relevant direction mainly lies in the $\lambda_4$-$h^2$ plane.

We further calculate other critical exponents along the $\lambda_2$ direction. Among them, $\eta_b$ and $\eta_f$ are obtained according to Eqs.~(\ref{eta2}) and (\ref{eta1}) (see the Supplemental Material for details), $\nu$ is given by the inverse of the eigenvalue of the stability matrix along the $\lambda_2$ direction, and others are obtained via the scaling law. We show the results in Table \ref{tabexp}. For comparison, the critical exponents for the tricritical Ising and the chiral Ising universality class are also displayed therein. Apparently, the chiral tricritical universality class is a new universality class distinct from both tricritical Ising and chiral Ising universality classes. From Table \ref{tabexp}, one can find that $\nu$ in the chiral trcritical Ising universal class is smaller than that in the chiral Ising universality class. The reason can be traced back to the fluctuation effect, which is weaker in the chiral tricritical universality class because of the absence of the divergent length scale in the first-order phase transition side of the CTP.

{\it Chiral Heisenberg tricritical point.---}
We extend the discussion of the chiral tricritical Ising universality to the chiral tricritcal point involving vector bosons, namely, the chiral tricritical Heisenberg point. The Lagrangian is given by
\bea
	\mathcal{L} &=& \bar \psi \gamma^\mu \partial_\mu  \psi+ \frac12 (\partial_\mu \vec \phi)^2+ h \vec \phi \cdot (\bar \psi  \vec s \psi) +  U(\rho),
\eea
where $\psi$ labels the fermionic operator the same as before, $\vec \phi$ refers to a three-component vector boson, and $\vec s$ is the Pauli matrix in the spin space. $U(\rho)$ is an analytical function of $\rho\equiv (\vec \phi)^2/2$, representing the potential of vector bosons.

Similar to the process in the previous model, we solve the RG flow equations and study the properties of the stability matrix near the fixed point (see the Supplemental Material). We find that the CTP also exists in the chiral Heisenberg model in which the two largest eigenvalues are positive, in contrast to the chiral Heisenberg fixed point, which has only one positive eigenvalue. We also calculate critical exponents and show the results in Table \ref{tabexp}.

{\it Discussion.---} The CTP studied in this Letter is a new fixed point located in an unexploited region of the parameter space. Given the theoretic prediction in this Letter, experimental and numerical studies are urgently called for. In particular, the critical exponents obtained above provide perceptible information for experiments. As hosted in the lattice model we construct, the CTP can manifest itself in various systems, such as graphene and gaphene-like materials \cite{firsov2005, neto2009,honerkamp2008,herbut2009,herbut2015,polikarpov2013,sorella2016,adam2015,scherer2017}. This provides broad candidates to detect the CTP. For instance, recent experiments with a controlled epitaxial graphene have visualized a chiral symmetry breaking phase~\cite{Gutierrez2016}. The CTP could possibly be realized therein by tuning the interaction between the boson modes. Moreover, the controllable magnetism, which has been realized in ultracold atom systems~\cite{Greif2013,Greif2015,Mazurenko2017}, particularly in the honeycomb lattice~\cite{Greif2015}, also provides promising platforms to achieve the CTP. 

{\it  Conclusion.---} In this Letter, we have constructed a microscopic model on honeycomb lattice with both fermions and Ising variables featuring a new CTP. From the proposed microscopic model, we find a phase transition between a Dirac semimetal and a CDW insulator that breaks sublattice symmetry as a function of external fields, and show the CTP residing between the first-order phase transition and the continuous phase transition. After extracting the effective action, we have employed the FRG method to explore the critical properties. We have found that this CTP is dictated by a new fixed point of the RG flow. The eigenvalues of the stability matrix at this fixed point indicate new critical phenomena which are qualitatively different from both the tricitical Ising universality class and the chiral Ising universality class. The experimental feasibility of realizing the CTP is also discussed.

{\it Acknowledgement. ---} We thank M. M. Scherer for helpful discussions. This work is supported in part by the NSFC under Grant No. 11474175 (S. Y., S.-K. J. and H. Y.) and by the MOST of China under Grant No. 2016YFA0301001 (H. Y.). S. Y. is supported in part by China Postdoctoral Science Foundation (Grant No. 2017M620035).

\begin{widetext}

\section{Supplemental Material}
\renewcommand{\theequation}{S\arabic{equation}}
\setcounter{equation}{0}
\renewcommand{\thefigure}{S\arabic{figure}}
\setcounter{figure}{0}
\renewcommand{\thetable}{S\arabic{table}}
\setcounter{table}{0}

\subsection{A. Mean field theory of tricritical Ising point without fermions}
The Ising Hamiltonian is given by
\bea
	H_b = J_1 \sum_{< ij >} \sigma_i^z \sigma_j^z - J_2 \sum_{\ll ij \gg} \sigma_i^z \sigma_j^z - H_\perp \sum_i \sigma_i^z - H_{\parallel} \sum_i \sigma_i^x.
\eea
In order to formulate the mean field theory, let's assume the vacuum expectation value (VEV) of Ising variable is given by $\langle \sigma_{A/B}^z \rangle = m \pm \phi $, where $A/B$ refer to two sublattices in honeycomb lattice and $m$ ($\phi$) refer to the uniform (staggered) VEV of Ising variable. Using the mean field decoupling,
\bea
	\sigma_i^z \sigma_j^z= \langle \sigma_i^z \rangle \sigma_j^z+  \sigma_i^z  \langle \sigma_j^z \rangle - \langle \sigma_i^z \rangle \langle \sigma_j^z\rangle,
\eea
we have
\bea
	\frac{H^\text{MF}_b}{M}= (J_- m- J_+ \phi - H_\perp) \sigma_A^z - H_{\parallel} \sigma_A^x+ (J_- m + J_+ \phi - H_\perp) \sigma_B^z - H_{\parallel} \sigma_B^x - J_- m^2 + J_+ \phi^2,
\eea
where $J_\pm= 3J_1 \pm 6 J_2$ and $M$ refers to the number of unit cells. Thus the MF energy per unit cell is given by
\bea
	\frac{E^\text{MF}_b}{M}= \pm \sqrt{(J_- m- J_+ \phi - H_\perp)^2+ H_{\parallel}^2} \pm \sqrt{(J_- m + J_+ \phi - H_\perp)^2+ H_{\parallel}^2}- J_- m^2+ J_+ \phi^2
\eea
For simplicity, we consider the case $J_-=0$ or equivalently $J_1=2J_2$, then the energy is independent of $m$ (note that the general features of the phase diagram does not depend on $J_-$ since we consider the condensation of $\phi$ field),
\bea
	\frac{E^\text{MF}_b}{M}= \pm \sqrt{(J_+ \phi + H_\perp)^2+ H_{\parallel}^2} \pm \sqrt{(J_+ \phi - H_\perp)^2+ H_{\parallel}^2}+ J_+ \phi^2.
\eea
The free energy per unit cell is $f_b= - \frac{1}{\beta N} \log \text{Tr}( e^{- \beta H^\text{MF}_b})$. At $T=0$, according to the free energy, the MF phase diagram is shown in Fig. \ref{phase2}. Near the phase boundary, assuming the free energy can be expanded as function of small $\phi$, we have $f_b= \sum_{n=0}^3 \lambda''_{2n} \phi^{2n}$ where
\bea
	\lambda''_0 = -2 \Delta,~~ \lambda''_2 = \Big(1- \frac{H_{\parallel}^2 J_+}{\Delta^3} \Big) J_+ ,~~ \lambda''_4 &=& \frac{(H_{\parallel}^2-4  H_\perp^2) H_{\parallel}^2J_+^4}{4 \Delta^7},~~ \lambda''_6 = \frac{(12 H_{\parallel}^4 H_\perp^2- H_{\parallel}^6 - 8 H_{\parallel}^2 H_\perp^4)J_+^6}{8 \Delta^{11}},
\eea
where $\Delta\equiv \sqrt{H_\perp^2 + H_{\parallel}^2}$.

\begin{figure}[b]
	\includegraphics[width=6cm]{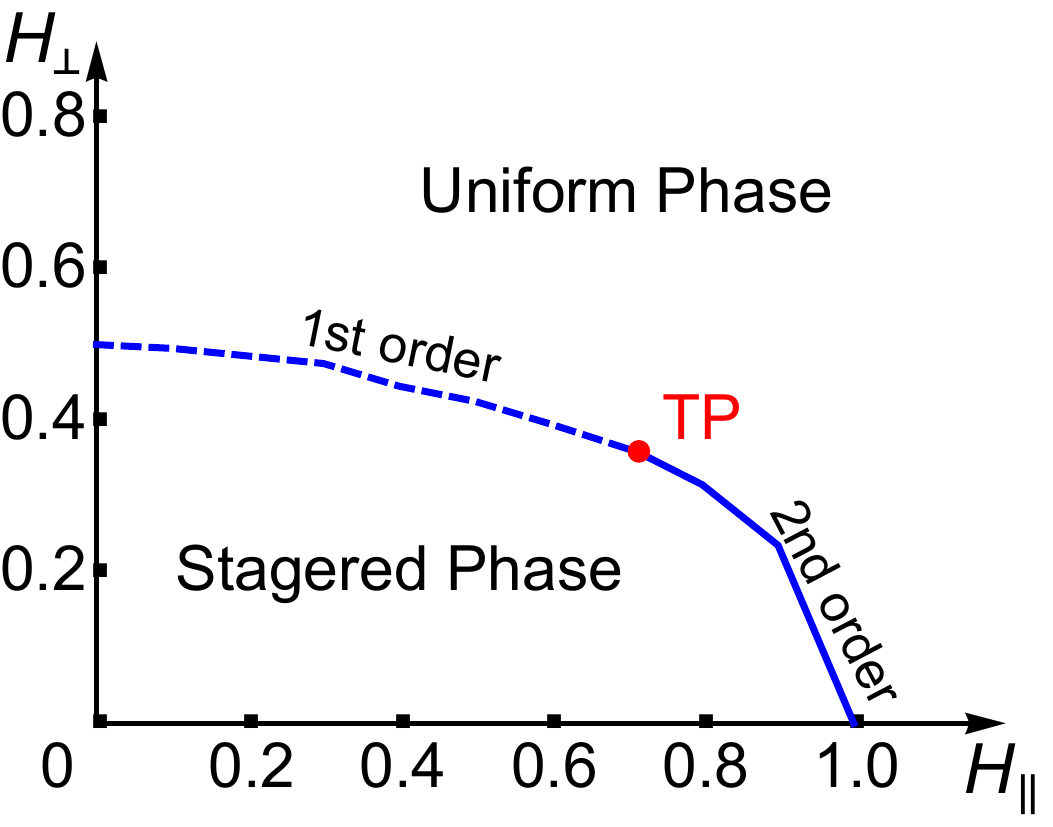}
	\caption{\label{phase2}Mean field phase diagram of $H_b$ at zero temperature as a function of Zeeman fields $H_\perp$ and $H_{\parallel}$ with $J_+=1, J_-=0$. The dashed and solid lines refer to first- and second-order phase transition between staggered phase and uniform phase. The red point denotes the tricritial point (TP).}
\end{figure}

\subsection{B. Mean field theory of chiral tricritical point}
Using Fourier transformation, the fermion Hamiltonian in momentum space reads
\bea
	H_f &=& \sum_k (c_{A,k}^\dag, c_{B,k}^\dag) \left( \ba{cccc} h \sigma^z_A  - \mu & -t(1+ e^{i k_2}+ e^{i(k_2-k_1)}) \\ -t(1+ e^{-i k_2}+ e^{-i(k_2-k_1)}) & h\sigma_B^z - \mu \ea \right) \left( \ba{cccc} c_{A,k} \\ c_{B,k} \ea \right)= \sum_k \tilde{c}^\dag_k \Big[ h m - \mu+ \varepsilon(k) \tau^z \Big] \tilde{c}_k, \nn \\
\eea
where in the last step we approximate $\sigma^z$ as its VEV and $\tilde{c}= (\tilde{c}_A, \tilde{c}_B)^T$ is the annihilation operator in  eigenwave basis and
\bea
	\varepsilon(k)= \sqrt{h^2 \phi^2+ t^2 [3+2(\cos k_1+ \cos k_2+ \cos(k_1-k_2))]},
\eea
here the lattice constant is set to be unity, and $\hat e_1=(1,0)$, $\hat e_2=(\frac12,\frac{\sqrt{3}}2)$ and $k_i= \hat e_i \cdot k$ for $i=1,2$. To perform the functional integral, we use the Lagrangian, $L_f = \sum_{k}  \tilde{c}_k^\dag [ -i\omega_n+  E(k, \tau)  ]\tilde{c}_k$ with $\omega_n \equiv \frac{(2n+1) \pi}{\beta}$ the Matsubara frequency. The free energy per site is
\bea
	f &=& - \frac{1}{\beta M} \log \text{Tr}( e^{- \beta ( H^\text{MF}_b+ H_f)})= - \frac{1}{\beta N} \log \sum_{\sigma}( e^{- \beta H_{MF}[\sigma]} \int D\tilde{c}^\dag D\tilde{c} e^{-S[\tilde{c},\tilde{c}^\dag]}) \\
	&=&- \frac{1}{\beta M} \log \sum_{\sigma} \Big[ e^{- \beta H^\text{MF}_b[\sigma]} \exp \sum_{n,k,\tau} \log \frac{1}{\beta}(-i\omega_n + E(k, \tau) )\Big] =  - \frac{1}{\beta M} \log \sum_{\sigma} \Big[ e^{- \beta H^\text{MF}_b[\sigma]} \exp \sum_{k,\tau} \log(1+ e^{-\beta E(k, \tau)})  \Big],  \nn \\
\eea
where $E(k, \tau)= h m - \mu+ \varepsilon(k) \tau^z$. At zero temperature, we have
\bea
f &\approx& -\sqrt{(J_+ \phi + H_\perp)^2+ H_{\parallel}^2} - \sqrt{(J_+ \phi - H_\perp)^2+ H_{\parallel}^2}+ J_+ \phi^2 + \frac{1}{M} \sum_{k,\tau} E(k, \tau) \theta(-E),
\eea
where $\theta$ is step function. In the disorder phase, we consider the neutral point $\rho= \frac{1}{\beta M} \frac{\partial}{\partial \mu} \log Z=0$. Thus the chemical potential is fixed to be $\mu=h m$ at zero temperature, and we have
\bea
 \frac{1}{M} \sum_{k,\tau} E(k, \tau) \theta(-E)= - \int \frac{d^2 k}{(2\pi)^2} \sqrt{h^2 \phi^2+ t^2 [3+2(\cos k_1+ \cos k_2+ \cos(k_1-k_2))]}.
\eea
From the free energy, we get the phase diagram shown in the main text. Assuming small $\phi$ near the phase boundary, we can expand the free energy as a function $\phi$. A subtlety arises here due to the coupling between Ising field and gapless Dirac fermions. Near $K=(\frac{4\pi}{3},0)$ and $K'=(-\frac{4\pi}{3},0)$ points, we make the approximation, $ 3+2(\cos k_1+ \cos k_2+ \cos(k_1-k_2)) \approx \frac34( k_x^2+k_y^2) $ for $k_x^2+k_y^2 \le \Lambda^2$ where $\Lambda$ is a cutoff, and
\bea
	&& -\int  \frac{d^2 k}{(2\pi)^2} \sqrt{h^2 \phi^2+ t^2 [3+2(\cos k_1+ \cos k_2+ \cos(k_1-k_2))]} \approx -2\int  \frac{d^2 k}{(2\pi)^2} \sqrt{h^2 \phi^2+ \frac34 t^2 k^2  } \\
	&=&-\frac{2(2\pi)}{(2\pi)^2} \int_\Lambda dk \sqrt{h^2 \phi^2+ \frac34 t^2 k^2  } \approx - \frac{t \Lambda^3}{2\sqrt{3}\pi}- \frac{h^2 \Lambda}{\sqrt{3}\pi t} \phi^2 + \frac{4 h^3}{9\pi t^2} |\phi|^3- \frac{h^4}{3\sqrt{3} \pi t^3 \Lambda} \phi^4 + \frac{2 h^6}{27 \sqrt{3} \pi t^5 \Lambda^3} \phi^6.
\eea
As a result the free energy near the phase transition is given by $f=\sum_{n=0}^3 \lambda_{2n}' \phi^{2n}+ \lambda_3 |\phi|^3$, where
\bea
	\lambda'_0 &=& -2 \Delta, ~~\lambda'_2=\left(1- \frac{H_{\parallel}^2 J_+}{\Delta^3} \right) J_+ - \frac{h^2 \Lambda}{\sqrt{3}\pi t},~~ \lambda'_4=\frac{(H_{\parallel}^2-4  H_\perp^2) H_{\parallel}^2J_+^4}{4 \Delta^7} - \frac{h^4}{3\sqrt{3} \pi t^3 \Lambda},  \\
	\lambda'_6 &=& \frac{(12 H_{\parallel}^4 H_\perp^2- H_{\parallel}^6 - 8 H_{\parallel}^2 H_\perp^4)J_+^6}{8 \Delta^{11}} +  \frac{2 h^6}{27 \sqrt{3} \pi t^5 \Lambda^3} , ~~\lambda_3= \frac{4 h^3}{9\pi t^2}.
\eea
Despite renormalization of the original coupling constants, coupling to gapless fermions gives rise to a nonanalyitical terms $|\phi|^3$.

\subsection{C. The effective Lagrangian near chiral tricritical point}
As mentioned in SM B, two Dirac cones locate at $K$ and $K'$. We now expand the free part of fermion Hamiltonian near the Dirac cones in the disordered phase, the low-energy Lagrangian of the Dirac fermion is given by
\bea
	\mathcal{L}_f &=& \psi^\dag [-i \omega+ v_F (k_x \tau^x \mu^z- k_y \tau^y)]\psi,
\eea
where $\psi= (\psi_{K,A}, \psi_{K,B}, \psi_{K',A}, \psi_{K',B})^T$ is fermion operator. $\omega$ is frequency while momentum $\vec k$ are measured from $K$ or $K'$ respectively. $\tau$ ($\mu$) refers to the Pauli matrix in sublattice (valley) space and $v_F= \frac{\sqrt{3}}{2}t a$, where $a$ is the lattice constant. The coupling between the Ising field $\phi$ and fermions are described by
\bea
	\mathcal{L}_{bf} = h \phi \psi^\dag \tau^z \psi,
\eea
where $h$ captures the coupling strength. The Lagrangian involving the Ising field is given by
\bea
	\mathcal{L}_f = \frac12 [(\partial_\tau \phi)^2+ v_B^2 (\vec \nabla \phi)^2] + U(\rho),
\eea
where $v_B$ is boson velocity and $U(\rho)$ is an analytical function of $\rho \equiv \phi^2$. Note that odd power of potential in the Ising field vanishes due to Ising symmetry. The expansion of potential near criticality is denoted $U(\rho)= \sum_i \lambda_{2i} \rho^i$.

It is well known that Lorentz symmetry will emerge at criticality, in the other word, the boson velocity and fermion velocity will flow to the same value in the IR. Thus, we set $v_F=v_B=1$ for simplicity. To further simplify the notation, we define $\bar \psi = \psi^\dag \gamma_0 $ and $\gamma^0= \tau^z$, $\gamma^1= \tau^y \mu^z$, $\gamma^2= \tau^x$, and $\partial_\mu= (\partial_\tau,\vec \nabla)$,
\bea
\mathcal{L}= \bar \psi \gamma^\mu \partial_\mu \psi+ \frac12 (\partial_\mu \phi)^2+h \phi \bar \psi \psi +  U(\rho),
\eea
which gives the Lagrangian in the main text.

\begin{table}[h]
  \centering
  \caption{The largest three eigenvalues of the stability matrix for different universality classes with $N_f=2$ in 2+1 dimensions. The eigenvalues corresponding to the conventional tricritical point are exact because its upper critical dimension is three.}
    \begin{tabular}{c| c| c| c }
    \hline
    \hline
     Universality class & $\theta_1$   &  $\theta_2$  & $\theta_3$   \\
    \hline
    \hline
       Chiral tricritical Ising   &2.2972  & 0.9816  & $-0.8739$   \\
    \hline
    Tricritical Ising		& 2 & 1 & 1 \\
     \hline
      Chiral Ising	               &0.9819 & $-0.8723$   & $-1.0897$  \\
    \hline
    \hline
         Chiral tricritical Heisenberg   &2.4818  & 0.7829  & $-0.9364$    \\
      \hline
      	Chiral Heisenberg	             &0.7716 & $-0.9238$   & $-1.5235$  \\
\hline
    \end{tabular}%
  \label{tab1}%
\end{table}%

\subsection{D. Eigenvalues of the stability matrices at different fixed points}
To explore critical properties near different fixed points, we distill the stability matrix at each fixed point. We list the largest three eigenvalues of $S$ in Table \ref{tab1}. The critical exponent $\nu$ is obtained by using $\nu \equiv \frac1\theta$, where $\theta$ is the eigenvalue of the stability matrix in the $\lambda_2$ direction.

\subsection{E. Method to determine the critical exponents}
Here we show the method to determine anomalous dimensions used in the main text. We use the extrapolation method to obtain the critical exponents at $N\rightarrow\infty$, where $N$ denotes the truncation order of the boson potential. At first, we fit out the dependence of the exponents as a polynomial function of $1/N$. The intercept of this function is just the exponents at $N\rightarrow\infty$. Then we extrapolate the fixed point value of $\lambda_i$ at $N\rightarrow\infty$ in the similar way and substitute the result to the function of $\eta_{b/f}$. Comparing the results obtained by these two different approaches, we can check the reliability of our method.
\begin{figure}\label{s2}
\subfigure[]{
	\includegraphics[width=5cm]{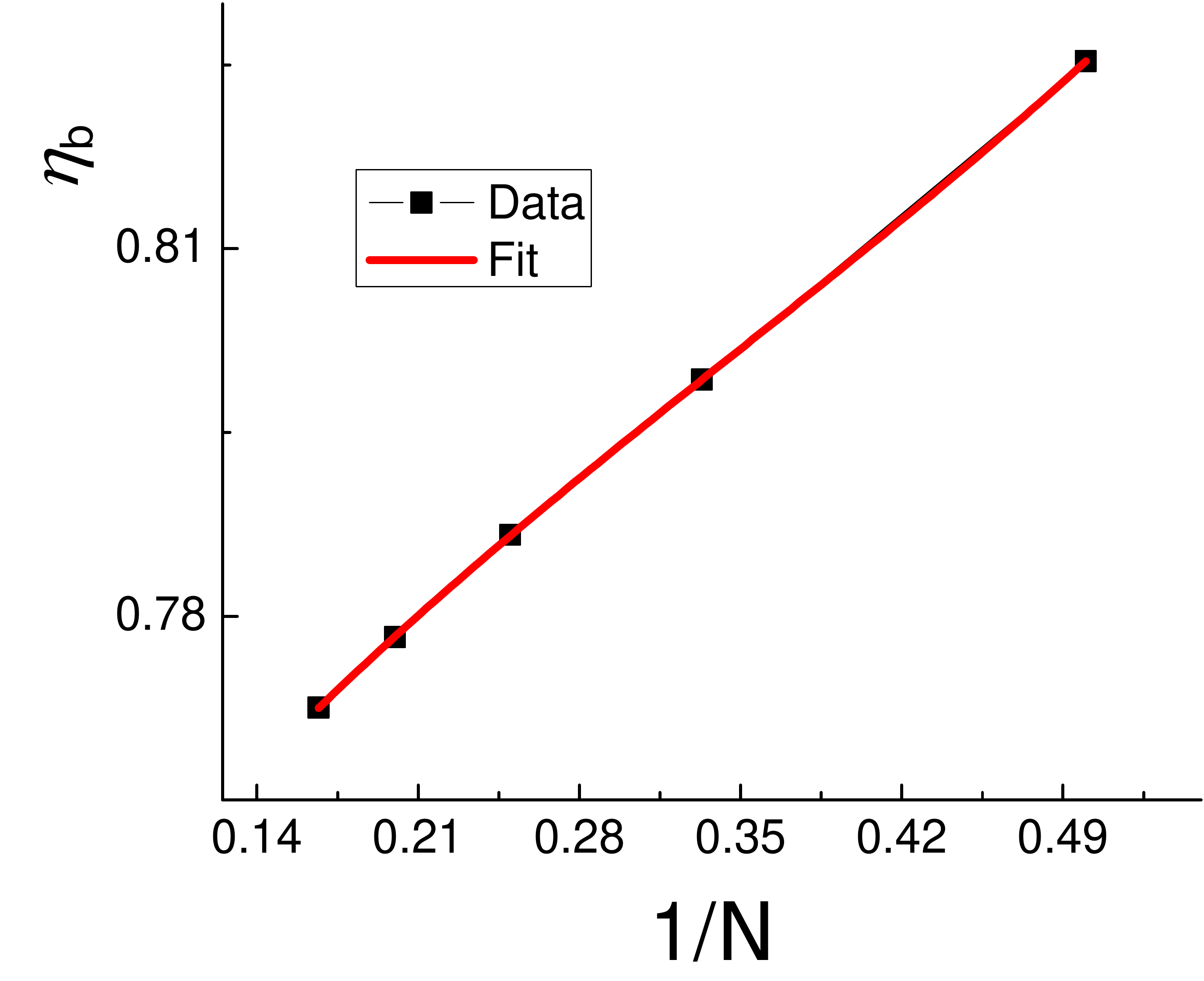}} ~~
\subfigure[]{\label{f}
	\includegraphics[width=6cm]{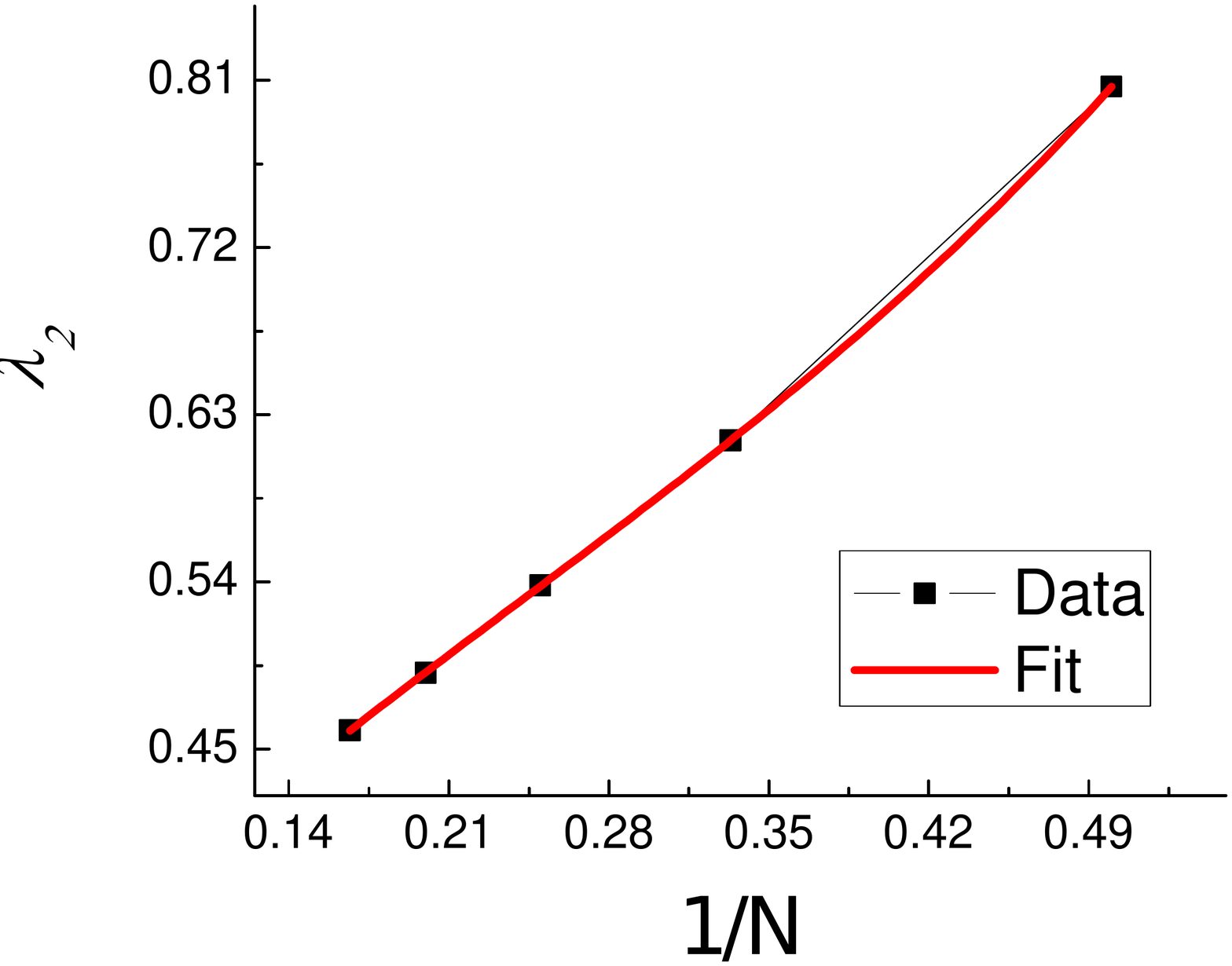}}
	\caption{\label{etafit} Third-order polynomial fit of the curve of $\eta_b$ and $\lambda_2$ versus $1/N$, where $N$ denotes the truncation order of the boson potential. The resulting intercept gives $\eta_b=0.7359$ and $\lambda_2=0.2763$ at $N\rightarrow\infty$.}
\end{figure}

The $N_f=2$ case in the chiral Ising model will be take as an example. In Fig. S2 we show the fitting of $\eta_b$ as a third-order polynomial function of $1/N$. From Fig. S2(a), we obtain $\eta_b=0.7359$. Similarly, we also obtain $\eta_f=0.0364$. Figure S2(b) shows the fitting of $\lambda_2$, which is necessary for calculating $\eta_b$ in LPA'. The intercept of the functions for $\lambda_2$ is $\lambda_2=0.2763$. Similarly, we obtain $h^2=2.2115$. By substituting them into Eqs. (11)-(14) in the main text, we obtain $\eta_b=0.7357$ and $\eta_f=0.0374$. Both values are consistent with those obtained by the direct fitting discussed above.

\end{widetext}
\end{document}